\documentclass[preprint,12pt]{elsarticle}
\usepackage[margin=2.5cm]{geometry}
\usepackage{amssymb}

\usepackage{subfiles}
\usepackage{tikz}
\usepackage{pgfplots}
\usepgfplotslibrary{colorbrewer}
\pgfplotsset{compat=newest}
\usetikzlibrary{positioning, calc, backgrounds, shapes.geometric, shapes.arrows, arrows.meta, shapes.symbols, decorations.text}
\usetikzlibrary{external}
\usepgfplotslibrary{colormaps}
\usepackage[most]{tcolorbox}
\usepackage[hidelinks]{hyperref}

\usepackage{mathrsfs}

\journal{arXiv}
\renewcommand{\ss}{\textsuperscript}
\newcommand{\F}{\mathbf{F}}
\newcommand{\Fdot}{\mathbf{\dot{F}}}
\renewcommand{\Finv}{\mathbf{F}^{-1}}
\newcommand{\FinvT}{\mathbf{F}^{-T}}
\newcommand{\Fe}{\mathbf{F}_e}

\newcommand{\Fedot}{\mathbf{\dot{F}}_e}

\renewcommand{\b}{\mathbf{b}}
\newcommand{\be}{\mathbf{b}_e}

\newcommand{\bedot}{\mathbf{\dot{b}}_e}
\newcommand{\tauneq}{\mathbf{\tau}_{NEQ}}
\newcommand{\eqn}[1]{(\ref{#1})}
\renewcommand{\S}{\mathbf{S}}
\newcommand{\C}{\mathbf{C}}

\newcommand{\Psieq}{\Psi_{EQ}}
\newcommand{\Psineq}{\Psi_{NEQ}}
\newcommand{\partiald}[2]{\frac{\partial #1}{\partial #2}}

\DeclareMathOperator*{\argmin}{arg\,min}
\newtheorem{remark}{Remark}[section]
\newcommand{\tr}{\mathrm{tr}}

\begin{document}

\begin{frontmatter}

\title{Data-driven anisotropic finite viscoelasticity using neural ordinary differential equations}
\author[inst1]{Vahidullah Tac}
\author[inst2]{Manuel Rausch}
\author[inst3]{Francisco Sahli Costabal}
\author[inst1,inst4]{Adrian Buganza Tepole}
\affiliation[inst1]{organization={Department of Mechanical Engineering, Purdue University}, 
            city={West Lafayette},
            state={IN},
            country={USA}}
\affiliation[inst2]{organization={Department of Aerospace Engineering and Engineering Mechanics, The University of Texas at Austin}, 
            city={Austin},
            state={TX},
            country={USA}}
\affiliation[inst3]{organization={Department of Mechanical Engineering, Pontifica Universidad Catolica de Chile}, 
            city={Santiago},
            country={Chile}}
\affiliation[inst4]{organization={Weldon School of Biomedical Engineering, Purdue University}, 
            city={West Lafayette},
            state={IN},
            country={USA}}

\begin{abstract}
We develop a fully data-driven model of anisotropic finite viscoelasticity using neural ordinary differential equations as building blocks. We replace the Helmholtz free energy function and the dissipation potential with data-driven functions that a priori satisfy physics-based constraints such as objectivity and the second law of thermodynamics. Our approach enables modeling viscoelastic behavior of materials under arbitrary loads in three-dimensions even with large deformations and large deviations from the thermodynamic equilibrium. The data-driven nature of the governing potentials endows the model with much needed flexibility in modeling the viscoelastic behavior of a wide class of materials. We train the model using stress-strain data from biological and synthetic materials including humain brain tissue, blood clots, natural rubber and human myocardium and show that the data-driven method outperforms traditional, closed-form models of viscoelasticity. 
\end{abstract}


\end{frontmatter}
\section{Introduction}

The traditional approach to modeling the mechanics of materials has been the use of expert-constructed closed-form constitutive equations. However, there are a large number of such models and no consensus on the best choice for any material. For example, Dal et al. \cite{dalPerformanceIsotropicHyperelastic2021} list 44 different hyperelastic constitutive material models for elastomers alone and, similarly, there are several examples of viscoelastic models for rubbers and polymers \cite{xiang2020review}.  
Furthermore, closed-form models  inherently restrict the type of behaviors that can be described, often rendering them incapable of accurately capturing the response of many materials. Both of these problems can be solved with the help of data-driven methods as has been demonstrated various times for the case of hyperelasticity \cite{tacDatadrivenTissueMechanics2022, tacDatadrivenModelingMechanical2022, liuGenericPhysicsinformedNeural2020, weberConstrainedNeuralNetwork2021}, but remains an ongoing area of investigation for dissipative phenomena such as viscoelasticity.

The past few years have witnessed a rapid adoption of data-driven methods in constitutive modeling of materials. 
Initially some studies used feed-forward fully-connected neural networks (FFNNs) to predict the derivatives of strain energy with respect to invariants of deformation. Liu et al. \cite{liuGenericPhysicsinformedNeural2020} enforced convexity of strain energy with respect to the elements of the Green strain tensor by adding loss terms that ensure the Hessian matrix of the strain energy is positive semi-definite. Other studies \cite{vlassisGeometricDeepLearning2020a, lengPredictingMechanicalProperties2021} used the monotonicity of stress as an alternative criterion to enforce the same class of convexity. However, polyconvexity is the more physically relevant criterion. We have previously enforced polyconvexity of the strain energy through the loss function \cite{tacDatadrivenModelingMechanical2022}. All of these studies enforced the desired constraint in the weak sense using specially designed loss functions that penalize deviations from convexity. Later, methods were developed using input convex neural networks (ICNNs) \cite{amosInputConvexNeural2017} and neural ordinary differential equations (NODEs) \cite{chenNeuralOrdinaryDifferential2019} to satisfy convexity conditions \textit{a priori} \cite{kleinPolyconvexAnisotropicHyperelasticity2022, tacDatadrivenTissueMechanics2022, CHEN2022103993, asadMechanicsInformedArtificial2022}. A parallel approach to constitutive modeling of hyperelasticity has been the automated discovery of constitutive laws from a large catalog of existing closed-form expressions \cite{linkaAutomatedModelDiscovery2022,linkaNewFamilyConstitutive2022,flaschelUnsupervisedDiscoveryInterpretable2021,thakolkaranNNEUCLIDDeeplearningHyperelasticity2022}. 


In comparison, adoption of data-driven methods in viscoelasticity has been slow. Some  forms of viscoelasticity models were developed for specific use cases, but no general, physics-informed, fully data-driven model of finite viscoelasticity has been developed so far, to the best of our knowledge. For example, Salahshoor and Ortiz \cite{salahshoorModelFreeDataDrivenViscoelasticity2022} took a model-free data-driven approach to viscoelasticity in the frequency domain to simulate wave propagation in viscoelastic solids. Wang et al. \cite{wangLongtimeIntegrationParametric2023} recently proposed a method of numerically approximating the solution of time-dependent ODEs and PDEs over long temporal domains using Physics Informed Deep Operator Neural Networks (PI DeepONets). The DeepONets are trained with a distribution of initial conditions and predict the solution of the ODE/PDEs in a short time interval. This solution is then used as the initial conditions for the next time interval. The authors of \cite{xuLearningViscoelasticityModels2021} use artificial neural networks to approximate the relationship between stress and strain increments in the discretized form of a viscoelastic constitutive law. And in \cite{chenRecurrentNeuralNetworks2021} the stress update algorithm is replaced entirely by recurrent neural networks (RNNs). However, the use of such approaches in large scale simulations would require additional safeguards because, in general, when the relationship between stress and strain is approximated using plain, unconstrained artificial neural networks (including the recurrent unit in an RNN) the positive dissipation of energy is not guaranteed, resulting in possible violations of the second law of Thermodynamics. An approach to achieve physics-informed viscoelasticity models is to represent the viscoelastic response as a Prony series with many terms and identify the most relevant parameters \cite{marino2022automated}, or to do sparse systems identification on a large library of models \cite{flaschel2022automated}. 

The lethargy in adoption of data driven methods in viscoelasticity can largely be attributed to two major challenges in developing fully data-driven, flexible, but physically realistic models of viscoelasticity: i) the difficulties associated with ensuring dissipation of energy is non-negative and, ii) high computational costs associated with optimizing the parameters of such models. 

\begin{figure*}[h!]
\centering
\includegraphics{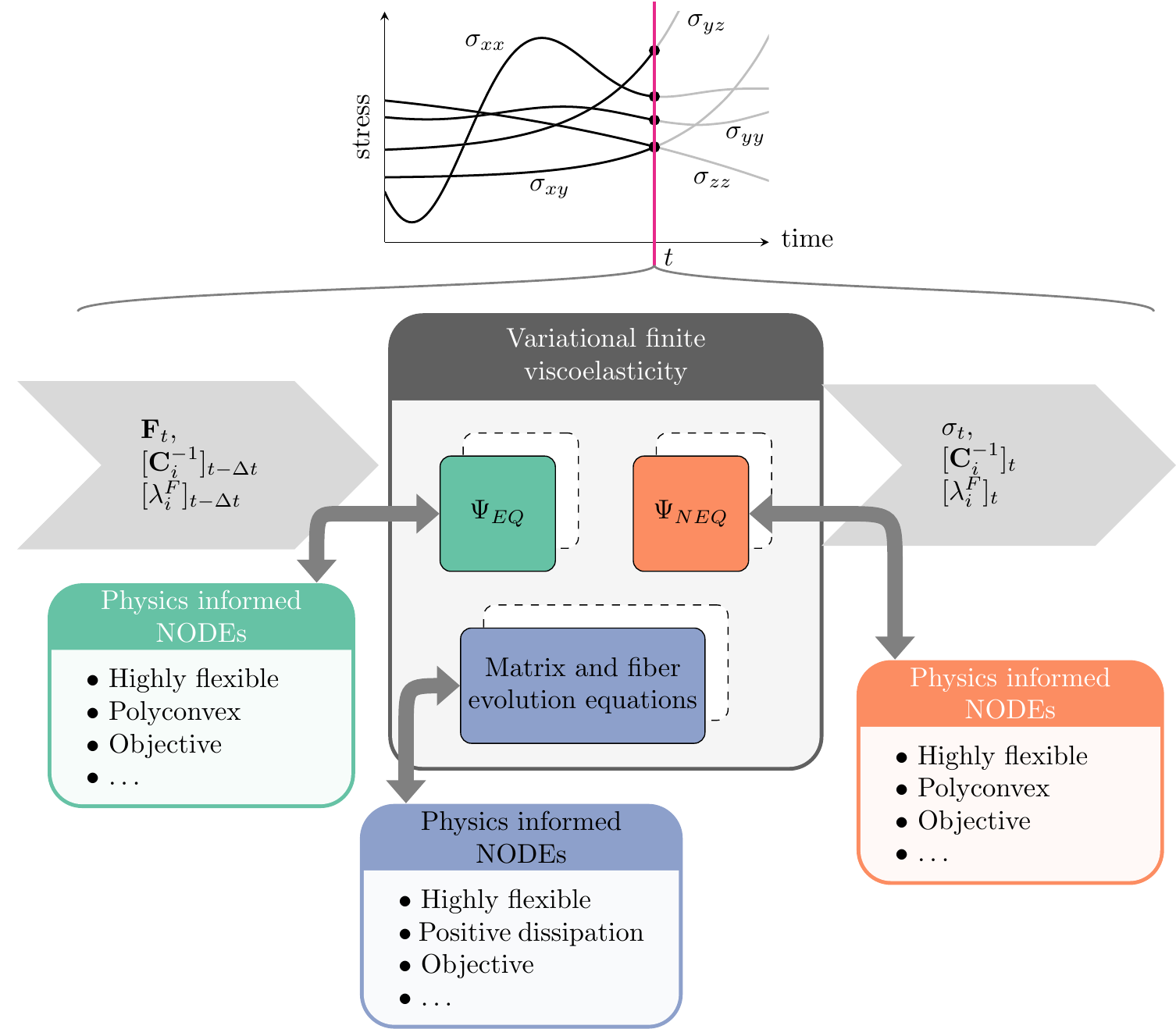}
\caption{Overview of the proposed method to construct data-driven constitutive models of finite viscoelasticity that automatically satisfy objectivity, polyconvexity and the dissipation inequality. The governing potentials $\Psieq$, $\Psineq$ and $\Phi$ are constructed using neural differential equations (NODEs) to satisfy physics-based constraints \textit{a priori}. The constitutive model can be used to compute stress histories for arbitrary deformation gradients and rates of deformation.}
\label{fig01} 
\end{figure*}

In this study we develop a fully data-driven model of anisotropic finite viscoelasticity using neural ordinary differential equations (NODEs). We adopt the framework of variational finite viscoelasticity from \cite{reeseTheoryFiniteViscoelasticity1998} as the mathematical foundation of the model and use the method of \cite{nguyenModelingAnisotropicFinitedeformation2007} to extend it to the anisotropic case. We propose NODE-based functions for the governing potentials Helmholtz free energy function and the \emph{creep potential} and a NODE-based evolution equation for the fiber stress. We show that some appropriate convexity conditions on the creep potential can guarantee that the dissipation of energy in the matrix is positive. We leverage prior knowledge accumulated in the field of hyperelasticity to design automatically convex data-driven creep potential functions using neural ordinary differential equations. The method provided in \cite{reeseTheoryFiniteViscoelasticity1998} for solving the evolution equation involves the use of a predictor-corrector algorithm, which is computationally costly and unstable when combined with gradient-based optimization algorithms, which are the mainstays of data-driven methods. We reformulate the problem as a system of ordinary differential equations (ODEs) thereby bypassing the iterative approach of the predictor-corrector algorithm. The Helmholtz free energy function is polyconvex with respect to the deformation gradient by construction \cite{tacDatadrivenTissueMechanics2022}. We train the model with experimental data from biological and synthetic materials including brain tissue samples, blood clots, natural rubber and mydium. A brief overview of this approach is shown in Fig. \ref{fig01}.

\section{Materials and Methods}
\subsection{Theory of finite viscoelasticity}
\label{theory_of_visco}
We assume that the material of interest consists of an isotropic matrix and a finite number of embedded fibers that endow the material with different mechanical behavior in certain directions, resulting in an overall anisotropic behavior. To model this phenomena, we use the parallel multiplicative decomposition of the deformation gradient into elastic and inelastic parts (denoted by subscripts $(\cdot)_e$ and $(\cdot)_i$, respectively) for both the matrix and the fibers (denoted by superscripts $(\cdot)^M$ and $(\cdot)^F$, respectively)
\begin{align}
\label{eq_split}
    \F = \F_e^M \F_i^M = \F_e^F \F_i^F
\end{align}
We additively split the Helmholtz free energy function into isotropic matrix and anisotropic fiber parts, both of which are further additively split into equilibrium (EQ) and non-equilibrium (NEQ) portions 
\begin{align}
    \label{eq_Psidef}
    \Psi = \Psieq^M(\C) + \Psieq^F(\C) + \Psineq^M(\C_e^M) + \Psineq^F(\C_e^F)
\end{align}
where $\C = \F^T \F$ is the right Cauchy-Green deformation tensor and $\C_e^M = {\F_e^M}^T \F_e^M$ and $\C_e^F = {\F_e^F}^T \F_e^F$ are the corresponding elastic counterparts for matrix and fiber, respectively. The matrix free energy functions are isotropic by construction and only depend on the isotropic invariants of $\C$ and $\C_e$, whereas the fiber functions are anisotropic and depend explicitly on deformation pseudo-invariants associated with fiber directions $\mathbf{V}$ defined in the reference configuration. 

\subsection{Helmholtz free energy}
\label{sec_helmholtz}
The construction of the free energy function is based on 
the NODE aproach from our previous publication \cite{tacDatadrivenTissueMechanics2022}. 

\subsubsection{Matrix phase}
We express the two free energy functions $\Psi^M_{EQ}$ and $\Psi^M_{NEQ}$ as the sum of an isochoric and a volumetric part
\begin{align}
    \Psi^M_{EQ}(\C) &= \Psi_{EQ}^{iso}(\bar{\C}) + \Psi_{EQ}^{vol}(J)
    \\
    \Psi^M_{NEQ}(\C_e) &= \Psi_{NEQ}^{iso}(\bar{\C}_e^M) + \Psi_{NEQ}^{vol}(J_e^M)
\end{align}
where $J=\det \F, J_e^M=\det \Fe^M$ and $\bar{\C}$ and $\bar{\C}_e^M$ are the isochoric parts of $\C$ and $\C_e^M$, respectively
\begin{align}
    \bar{\C} &= J^{-2/3} \C
    \\
    \bar{\C}_e^M &= J_e^{-2/3} \C_e^M \,.
\end{align}
We require that the two free energy functions $\Psi^M_{EQ}$ and $\Psi^M_{NEQ}$ be objective and polyconvex in $\F$ and $\Fe^M$ respectively. To this end, we express the $\Psi_{EQ}^{iso}$ and $\Psi_{NEQ}^{iso}$ in terms of isotropic polyconvex invariants of $\hat{\C}$ and $\hat{\C}_e^M$ \cite{hartmannPolyconvexityGeneralizedPolynomialtype2003}, respectively, and adopt an additive split into two independent functions
\begin{align}
    \Psi_{EQ}^{iso}(\C) &= \Psi_{EQ}^{iso}(\bar{I}_1, \tilde{I}_2) = \Psi_{EQ,1}(\bar{I}_1) + \Psi_{EQ,2}(\tilde{I}_2)
    \\
    \Psi_{NEQ}^{iso}(\C_e^M) &= \Psi_{NEQ}^{iso}(\bar{I}_1^e, \tilde{I}_2^e) = \Psi_{NEQ,1}(\bar{I}_1^e) + \Psi_{NEQ,2}(\tilde{I}_2^e)
\end{align}
with
\begin{align}
    \bar{I}_1 &= \tr(\bar{\C})
    \\
    \tilde{I_2} &= \bar{I}_2^{3/2} - 3\sqrt{3} \text{ with } \bar{I}_2 = \frac{1}{2}\left( \bar{I}_1^2- \tr(\bar{\C}^2) \right)
    \\
    \bar{I}_1^e &= \tr((\bar{\C}_e^M))
    \\
    \tilde{I_2^e} &= (\bar{I}_2^e)^{3/2} - 3\sqrt{3} \text{ with } \bar{I}_2^e = \frac{1}{2}\left( (\bar{I}_1^e)^2- \tr(\bar{\C}_e^M)^2 \right) \, .
\end{align}


Then, to guarantee polyconvexity of $\Psi_{EQ}$ and $\Psi_{NEQ}$ it is sufficient to require that the functions $\Psi_{EQ,1}, \Psi_{EQ,2}, \Psi_{EQ}^{vol}, \Psi_{NEQ,1}, \cdots$ be convex and non-decreasing. Or equivalently, it is sufficient for 
$\partial \Psi_{EQ,1}/\partial\bar{I}_1$, $\partial \Psi_{EQ,2}/\partial\tilde{I}_2$ be monotonic and non-negative. We use NODEs to approximate the functions $\partial \Psi_{EQ,1}/\partial\bar{I}_1$, $\partial \Psi_{EQ,2}/\partial\tilde{I}_2$, such that they are monotonic and non-negative as explained in the Section \ref{NODEs}.

\subsubsection{Fiber phase}
The Helmholtz free energy function of the fiber phase can be expressed in terms of anisotropic pseudo-invariants of the deformation. In this study we use the following form for $\Psieq^F$ and $\Psineq^F$:
\begin{align}
    \Psieq^F(\C) &= \Psieq^F(I_4)
    \\
    \label{eq_psineq_F}
    \Psineq^F(\C_e^F) &= \Psineq^F({I^F_{4e}})
\end{align}
with
\begin{align}
\label{eq_I4_I4e}
    I_4 &= \C:\mathbf{M}
    \\
    \label{eq_def_I4e_F}
    I^F_{4e} &= \frac{\C:\mathbf{M}}{\C_i^F:\mathbf{M}} 
\end{align}
where $\mathbf{M} = \mathbf{V}\otimes\mathbf{V}$ is the structure tensor defined in terms of material fiber direction vector $\mathbf{V}$. 
As in the matrix case, we use NODEs to approximate the derivative functions 
$\partial \Psieq^F/\partial I_4$, $\partial \Psineq^F/\partial I_{4e}^F$.

\subsection{Dissipation inequality}
The constitutive equations have to satisfy the entropy inequality, which is equal to the \emph{internal dissipation inequality} in the case of isothermal processes
\begin{align}
    \label{eq_diss_ineq}
    \frac{1}{2}\S : \dot{\C} + \dot{\Psi} \geq 0
\end{align}
with $\S = 2\partial {\Psi}/\partial{\C}$ the second Piola-Kirchhoff stress. It can be shown that substitution of \eqref{eq_Psidef} into \eqref{eq_diss_ineq} leads to \cite{nguyenModelingAnisotropicFinitedeformation2007}
\begin{align}
    \label{eq_diss_ineq_expanded}
    -2\partiald{\Psi}{\C_i^M}: \frac{1}{2} \dot{\C}_i^M -2\partiald{\Psi}{\C_i^F}: \frac{1}{2} \dot{\C}_i^F \geq 0 \, .
\end{align}

The two terms in \eqref{eq_diss_ineq_expanded} correspond to the dissipation of energy in the matrix and the fiber. We require that the two terms be non-negative individually. This corresponds to the assumption that the dissipation mechanisms in the matrix and the fiber are independent of each other, resulting in two separate criteria 
\begin{align}
    \label{eq_diss_ineq_matrix}
    -2\partiald{\Psi}{\C_i^M}: \frac{1}{2} \dot{\C}_i^M &\geq 0
    \\
    \label{eq_diss_ineq_fiber}
    -2\partiald{\Psi}{\C_i^F}: \frac{1}{2} \dot{\C}_i^F &\geq 0 \, .
\end{align}

\subsection{Evolution law for the matrix phase}
The inequality of \eqref{eq_diss_ineq_matrix} can be more conveniently written in spatial form as \cite{reeseTheoryFiniteViscoelasticity1998}
\begin{align}
    \label{eq_diss_ineq_final}
    -\tauneq^M : \frac{1}{2} (\mathscr{L} \be^M) ({\be^M})^{-1} \geq 0
\end{align}
where $\tauneq^M = 2\Fe^M (\partial {\Psineq^M}/\partial {\C_e^M}) \Fe^M $ is the Kirchhoff stress of the non-equilibrium part of the deformation, $\mathscr{L}(\cdot)$ denotes the Lie derivative and $\be^M = \Fe^M {\Fe^M}^T$ is the left Cauchy-Green deformation tensor of the elastic part of the matrix deformation. 
Eq. (\ref{eq_diss_ineq_final}) implies the need of a constitutive equation that specifies the time evolution of $\mathbf{b}_e^M$. One option is to introduce a scalar valued function $\Phi$, known as the \emph{creep potential}, which defines the evolution equation for the matrix phase

\begin{align}
    \label{eq_lie_to_phi}
    -\frac{1}{2} (\mathscr{L} \be^M) {\be^M}^{-1} = \partiald{\Phi}{\tauneq^M} \,.
\end{align}
Substituting \eqref{eq_lie_to_phi} to \eqref{eq_diss_ineq_final} and carrying out the double contraction results in
\begin{align}
    \sum_{i=1}^3\sum_{j=1}^3(\tauneq^M)_{ij} \partiald{\Phi}{(\tauneq^M)_{ij}} \geq 0
\end{align}
We require that this inequality be satisfied for each $(\tauneq^M)_{ij}$ independently, i.e.,
\begin{align}
    (\tauneq^M)_{ij} \partiald{\Phi}{(\tauneq^M)_{ij}} &\geq 0 \text{ (no sum) }\,\, \forall i,j \in \{1,2,3\}
    \\
    \implies \partiald{\Phi}{(\tauneq^M)_{ij}} &\geq 0 \text{ when } (\tauneq^M)_{ij} \geq 0 \,\, \text{ and }
    \\
     \partiald{\Phi}{(\tauneq^M)_{ij}} &\leq 0 \text{ when } (\tauneq^M)_{ij} \leq 0
\end{align}
Thus, it can be seen that to satisfy the dissipation inequality \eqref{eq_diss_ineq} it is sufficient to ensure 
\begin{enumerate}
    \item $\Phi$ is a convex function of  $\tauneq^M$
    \item $\argmin_{\tauneq^M} \Phi= \mathbf{0}$ (i.e., $\Phi$ attains its global minimum at $\tauneq^M = \mathbf{0}$)
\end{enumerate}
but the principle of objectivity of stress introduces a third requirement
\begin{enumerate}
  \item[3.] $\Phi$ is an objective function of $\tauneq$, i.e. $\Phi(\mathbf{R} \tauneq^M \mathbf{R}^T) = \Phi(\tauneq^M) $ for all proper orthogonal tensors $\mathbf{R}$
\end{enumerate}

\subsubsection{Convex and objective creep potentials}
\label{sec_conv_obj_creep_pot}
We are interested in finding convex and objective functions $\Phi: \mathbb{S}\mathrm{ym}^{3\times3} \to \mathbb{R}$ such that $\argmin_{\tauneq^M} \Phi= \mathbf{0}$. Here $\mathbb{S}\mathrm{ym}^{3\times3}$ denotes the space of symmetric $3\times3$ matrices. 
One approach to this problem is to make $\Phi$ a function of  the principal invariants or the eigenvalues of $\tauneq^M$. 
\begin{align}
    \label{eq_phi_inv}
    \Phi(\tauneq^M) = \hat{\Phi}(I_1^\tau, I_2^\tau, I_3^\tau, \tau_1, \tau_2, \tau_3)
\end{align}
where the principal invariants are denoted $I_i^\tau$, and the eigenvalues are $\tau_i$.

From the fundamentals of convex functions of multiple variables \cite{boydConvexOptimization2004} we know that a composition of functions $f=h(\mathbf{g}(\mathbf{x})) = h(g_1(\mathbf{x}), g_2(\mathbf{x}), \cdots)$ is convex in $\mathbf{x}$ if 
\begin{enumerate}
    \item $g_i: \mathbb{R}^n \to \mathbb{R}, i=1,2,\cdots k$ are convex in each argument
    \item $h\phantom{_i}: \mathbb{R}^k \to \mathbb{R}$ is convex and non-decreasing in each argument
\end{enumerate}

This suggests the following form for \eqref{eq_phi_inv}
\begin{align}
    \label{eq_phi_g}
    \Phi(\tauneq^M) = \hat{\Phi}(g_1(I_1^\tau, I_2^\tau, \cdots), g_2(I_1^\tau, I_2^\tau, \cdots), \cdots)
\end{align}
where $\hat{\Phi}$ is convex and non-decreasing and $g_i$ functions of invariants of $\tauneq$ such that $g_i$ is convex in $\tauneq^M$. Some suitable examples of such functions are:
\begin{itemize}
    \item The first principal invariant, $I_1^\tau = \tr(\tauneq^M)$ and its square $(I_1^\tau)^2$
    \item $(I_1^\tau)^2-\alpha I_2^\tau$ where $0\leq \alpha \leq 3$ and $I_2^\tau = \tr(\mathrm{cof}(\tauneq^M))$ is the second principal invariant of $\tauneq$
    \item The largest eigenvalue $\tau_1$ of $\tauneq^M$
    \item $-\tau_3$ where $\tau_3$ is the smallest eigenvalue of $\tauneq^M$ 
    \item Sum of $k$ largest eigenvalues
\end{itemize}
$I_1^\tau = (\tauneq^M)_{11} + (\tauneq^M)_{22} + (\tauneq^M)_{33}$ is linear in elements of $\tauneq$ and therefore its convexity is obvious. Proofs of convexity of $(I_1^\tau)^2$ and $(I_1^\tau)^2-\alpha I_2^\tau$ are given in \ref{ap_proof_convex}. For the proofs of the other functions given here, or more examples of convex functions the readers are referred to \cite{boydConvexOptimization2004}. 

In this study we adopt an additive form for $\hat{\Phi}$ and use the first three of the functions listed above, i.e. $g_1 = I_1^\tau, g_2 = (I_1^\tau)^2$ and $g_3 = (I_1^\tau)^2-3 I_2^\tau$. With this, \eqref{eq_phi_g} becomes
\begin{align}
    \Phi(\tauneq^M) = \hat{\Phi}_1(I_1^\tau) + \hat{\Phi}_2((I_1^\tau)^2) + \hat{\Phi}_3((I_1^\tau)^2- 3I_2^\tau) \,.
\end{align}
This form of $\Phi$ is sufficient for most practical applications. In fact it can be shown that the creep potential used in \cite{reeseTheoryFiniteViscoelasticity1998} consists of just a linear combination of $g_2 = (I_1^\tau)^2$ and $g_3 = (I_1^\tau)^2- 3I_2^\tau$, which is a special case of the class of creep potentials we can capture with NODEs.

Finally, for the requirement on the global minimum we have
\begin{align}
    \argmin_{\tauneq^M}\Phi(\tauneq^M) = \argmin_{\tauneq^M} \left( \hat{\Phi}_1(I_1^\tau) + \hat{\Phi}_2((I_1^\tau)^2) + \hat{\Phi}_3((I_1^\tau)^2- 3I_2^\tau) \right) = \mathbf{0}
\end{align}
we require that the functions $\hat{\Phi}_i$ individually attain their minima at $\tauneq^M = \mathbf{0}$, i.e.
\begin{align}
    &\argmin_{\tauneq^M} \hat{\Phi}_1\left(I_1^\tau(\tauneq^M)\right) = \mathbf{0}
    \\
    &\argmin_{\tauneq^M} \hat{\Phi}_2\left(\left(I_1^\tau(\tauneq^M)\right)^2\right) = \mathbf{0}
    \\
    &\argmin_{\tauneq^M} \hat{\Phi}_3\left((I_1^\tau(\tauneq^M))^2- 3I_2^\tau(\tauneq^M)\right) = \mathbf{0}
\end{align}
or, considering the fact that $I_1^\tau = (I_1^\tau)^2 = (I_1^\tau)^2- 3I_2^\tau = 0$ when $\tauneq^M = \mathbf{0}$, this condition can alternatively be stated as
\begin{equation}
    \argmin_{g_i} \hat{\Phi}_1(g_i) = 0\, ,\; i=\{1,2,3\} \, .
\end{equation}
These requirements can be equivalently stated in terms of the derivatives of $\hat{\Phi}_i$
\begin{align}
    \label{eq_req1}
    \hat{\Phi}_i(g_i) \text{ convex } &\Longleftrightarrow \frac{d\hat{\Phi}_i}{dg_i} \text{ monotonically increasing}
    \\
    \label{eq_req2}
    \hat{\Phi}_i(g_i) \text{ non-decreasing } &\Longleftrightarrow \frac{d\hat{\Phi}_i}{dg_i} \text{ non-negative}
    \\
    \label{eq_req3}
    \argmin_{g_i} \hat{\Phi}_i= 0 &\Longleftrightarrow \left.\frac{d\hat{\Phi}_i}{dg_i} \right|_{g_i=0} = 0
\end{align}
We use NODEs to approximate the derivative functions ${d\hat{\Phi}_i}/{dg_i}$, such that these three requirements are satisfied \textit{a priori}, see also Section~\ref{NODEs}. At this point we need to make two remarks
\begin{remark}
    $\Phi_1$ is not in the form of a composition of functions. Rather, it is linear in the elements of $\tauneq$, i.e., $\hat{\Phi}_1(I_1^\tau) = \hat{\Phi}_1((\tauneq^M)_{11} + (\tauneq^M)_{22} + (\tauneq^M)_{33})$. As such, $\hat{\Phi}_1$ does not need to be a non-decreasing function of its argument. In other words, only requirements \eqref{eq_req1} and \eqref{eq_req3} need to be satisfied for $\Phi_1$.
\end{remark}
\begin{remark}
    Notice that the argument of $\hat{\Phi}_2$ is non-negative. The argument of $\hat{\Phi}_3$ can be expressed in terms the eigenvalues of $\tauneq$
    \begin{align}
        (I_1^\tau)^2- 3I_2^\tau = \tau_1^2 + \tau_2^2 + \tau_3^2 - \tau_1\tau_2 -\tau_1\tau_3 - \tau_2\tau_3
    \end{align}
    which shows that it too is non-negative. This will be important in Section \ref{NODEs} where we will construct NODEs such that non-negative inputs lead to non-negative outputs thereby fulfilling the requirement \eqref{eq_req2}.
\end{remark}

\subsection{Evolution law for the fiber phase}
Substituting the free energy function of \eqref{eq_psineq_F} into the dissipation inequality of \eqref{eq_diss_ineq_fiber} yields
\begin{align}
    \label{eq_def_tau_neq_f}
    \underbrace{2{I^F_{4e}}\partiald{\Psineq^F}{{I^F_{4e}}}}_{\tau_{NEQ}^F} \frac{\mathbf{M}}{\C_i^F : \mathbf{M}} : \frac{1}{2} \dot{\C}_i^F \geq 0
\end{align}
where $\tau_{NEQ}^F \in \mathbb{R}$ denotes the fiber stress. Note that the double contraction in this equation corresponds to the inelastic part of the stretch in the fiber
\begin{align}
\label{eq_lam_iF}
    ({\lambda_i^F})^2 = I_{4i}^F= \C_i^F : \mathbf{M} \, .
\end{align}
With Eq. (\ref{eq_lam_iF}), the dissipation inequality can be written more concisely as
\begin{align}
\label{eq_diss_aniso}
    \tau_{NEQ}^F \left( \frac{\dot{\lambda}_i^F}{\lambda_i^F} \right) \geq 0 \, .
\end{align}

Eq. (\ref{eq_diss_aniso}) requires the specification of a constitutive equation for the evolution of $\lambda_i^F$. The model has to satisfy $\mathrm{sign}(\dot{\lambda}_i^F/\lambda_i^F) = \mathrm{sign} (\tau_{NEQ}^F)$. In particular, any monotonically increasing function passing through $(0,0)$ will satisfy this. In light of this, we propose the following form of the evolution equation for the fiber

\begin{align}
    \label{eq_fiber_evol_node}
    \left( \frac{\dot{\lambda}_i^F}{\lambda_i^F} \right) = \mathcal{N}(\tau_{NEQ}^F)
\end{align}
where $\mathcal{N}$ denotes a NODE. As we show in Section \ref{NODEs}, a NODE is guaranteed to be monotonic and it can be constructed to pass through the origin.

Note that Eq. (\ref{eq_fiber_evol_node}) is a scalar equation that should be enough to specify the evolution of $\C_i^F,\C_e^F$. One way of closing the system of question is to specify the tensor $\C_i^F$ from $\lambda_i^F$, for example 

\begin{equation}
\C_i^F = (\lambda_i^F)^2 \mathbf{M} \, .
\end{equation}

However, the evolution of the fiber viscoelastic branch can be done without the explicit multiplicative split of the deformation gradient (\ref{eq_split}). Instead, only the scalar multiplicative split

\begin{equation}
\label{eq_scalar_split_F}
    I_{4}^F = I_{4e}^F I_{4i}^F
\end{equation}

is needed. Then, eqs. (\ref{eq_def_tau_neq_f}), (\ref{eq_fiber_evol_node}) and (\ref{eq_scalar_split_F}) fully specify the evolution of the fiber viscoelastic branch without computing $\C_i^F$, $\C_e^F$. To close, the contribution to the stress tensor  is

\begin{align}
    \mathbf{\sigma}_{NEQ}^F = \frac{1}{J} \F \S_{NEQ}^F \F^T = \frac{2}{J} \F \partiald{\Psi_{NEQ}^F}{\C} \F^T = \frac{2}{J} \F \partiald{\Psi_{NEQ}^F}{I_{4e}^F} \partiald{I_{4e}^F}{\C} \F^T
\end{align}

which reduces to 
\begin{align}
\label{eq_sigma_neq_F}
    \mathbf{\sigma}_{NEQ}^F = \frac{1}{J}\tau_{NEQ}^F \frac{\F \mathbf{M} \F^T}{\C :\mathbf{M}} \, .
\end{align}

\subsection{Neural Ordinary Differential Equations}
\label{NODEs}
A NODE is a machine learning framework that generalizes some classical models such as recurrent neural networks. We use NODEs to construct monotonic and nonnegative relationships between inputs $\mathbf{x}$ and outputs $\mathbf{y}$. The key idea behind NODEs is the use of an ODE defined on pseudotime $\omega$ to compose continuous transformations of a hidden state $\mathbf{h}(\omega)$
\begin{align}
    \label{eq_ODE}
    \frac{d\mathbf{h}(\omega)}{d\omega} = f(\mathbf{h}(\omega), \omega, \mathbf{\theta})
\end{align}
where the function $f(\cdot, \cdot, \theta)$ is given by a fully connected feed-forward neural network with parameters $\mathbf{\theta}$. Given the initial value $\mathbf{h}(0)$ of the hidden state, the output of the model can be obtained by integrating the ODE up to a predetermined pseudotime, say $\omega=1$
\begin{align}
    \mathbf{h}(1) = \mathbf{h}(0) + \int_0^1f(\mathbf{h}(\omega), \omega, \mathbf{\theta})dt
\end{align}
We assign the input to the initial condition $\mathbf{h}(0) = \mathbf{x}$ and the final state to the output $\mathbf{y} = \mathbf{h}(1)$.

From the fundamentals of ODEs we know that no two trajectories of an ODE intersect, because the contrary would indicate that the right hand side $f(\cdot)$ will map the two trajectories to different direction at the point of the intersection, which contravenes the definition of a function. In the 1-dimensional case, this indicates that for any two trajectories $h_1(t)$ and $h_2(t)$
\begin{align*}
    h_1(0) \geq h_2(0) \Longleftrightarrow h_1(1) \geq h_2(1)
    \\
    h_1(0) \leq h_2(0) \Longleftrightarrow h_1(1) \leq h_2(1)
\end{align*}
or more succinctly
\begin{align*}
    (h_2(1)-h_1(1))(h_2(0)-h_1(0)) &\geq 0
    \\
    \implies (y_2-y_1)(x_2-x_1) &\geq 0
\end{align*}
which shows that the input-output map of a NODE is monotonic. 

As mentioned previously, the right hand side of the ODE in \eqref{eq_ODE} is given by a neural network 
\begin{equation}
    f(x) = \mathbf{W}_n h(...\mathbf{W}_2 h(\mathbf{W}_1x + \mathbf{b}_1) + \mathbf{b}_2)) + \mathbf{b}_n
\end{equation}
where $\mathbf{W}_i$ and $\mathbf{b}_i$ are the weights and biases of the $i$-th layer of the neural network. Removing the biases from the neural network results in a new feed-forward operation consisting of a series of multiplications that map $x=0$ to $f(x)=0$ provided the activation function satisfies $h(0)=0$,
\begin{equation}
    \label{eq_f_nobias}
    f(x) = \mathbf{W}_n h(...\mathbf{W}_2 h(\mathbf{W}_1x))) \,.
\end{equation}
Together with the monotonicity of NODEs, an architecture such as \ref{eq_f_nobias} guarantees that the outputs of the NODEs are non-negative and that the NODEs attain their global minima at $x=0$, i.e.,
\begin{align}
    f(x) = 0 \quad \implies y = 0 \quad  \text{ when } \quad x=0 
    \\
    f(x) > 0 \quad \implies y > 0 \quad \text{ when } \quad x>0 
\end{align}
While this form of NODEs is suitable, removing all the biases is not necessary. In application we use neural networks of the form
\begin{equation}
    \label{eq_f_posbias}
    f(x) = \mathbf{W}_n h(...\mathbf{W}_2 h(\mathbf{W}_1x))) + \exp (\mathbf{b}_n)
\end{equation}
This form of $f$ is identical to the one given in \eqref{eq_f_nobias} with the exception that the output of the neural network is shifted by a positive value, which adds more flexibility to the neural network. However, all the desirable qualities of NODEs that we have achieved so far are preserved. 

\begin{figure*}[h!]
\centering
\includegraphics{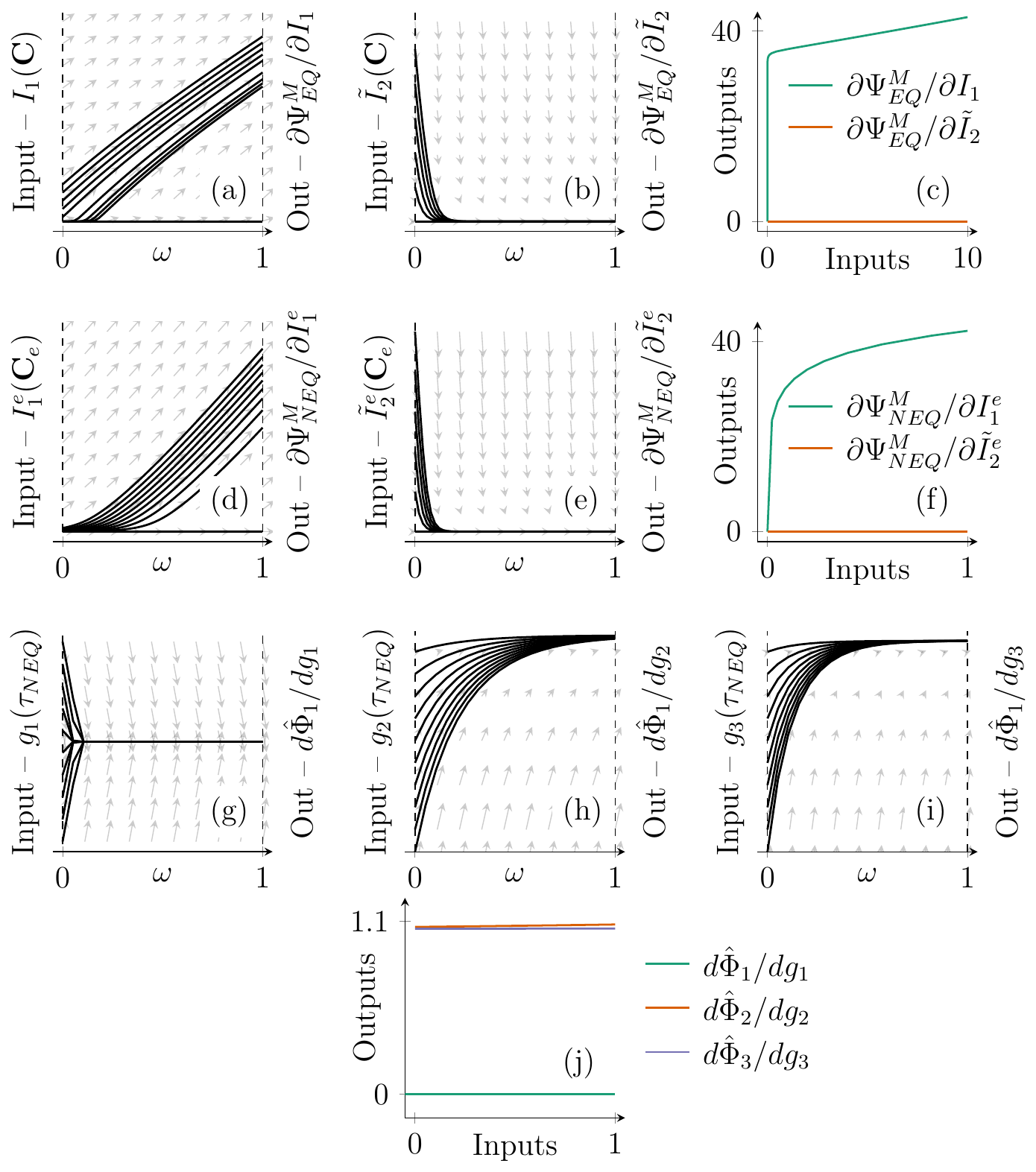}
\caption{Example of ODE trajectories generated by NODEs. Trajectories of the ODEs for $\partial {\Psieq^M}/\partial {I_1}$ (a) and $\partial {\Psieq^M}/ \partial{\tilde{I}_2}$ (b). The NODE defines the underlying vector field shown in grey, which generates the observed trajectories. The corresponding input-output map is monotonic (c). Trajectories of the ODEs for $\partial {\Psineq^M}/\partial {I_1^e}$ (d) and $\partial {\Psineq^M}/\partial {\tilde{I}_2^e}$ (e) and the resulting input-output maps (f) have a similar response to the equilibrium branch. Trajectories of the ODEs for $d \hat{\Phi}_1/d g_1$ (g), $d \hat{\Phi}_2/d g_2$ (h) and $d \hat{\Phi}_3/d g_3$ (i) with $g_i$ as defined in Section \ref{sec_conv_obj_creep_pot} and the resulting input-output maps (j). In this case the domain of the inputs includes negative non-equilibrium stresses and the outputs are convex functions with minimum at vanishing non-equilibrium stress. }
\label{fig02} 
\end{figure*}

In summary, we have shown that the input-output map of a NODE is monotonic. When neural networks of the form \eqref{eq_f_posbias} are used, non-negative inputs result in non-negative outputs. And finally, the NODEs attain their global minima at $x=0$. We use NODEs to approximate the functions $\partial \Psi_{EQ}/\partial I_1, \partial \Psi_{EQ}/\partial I_2, \partial \Psi_{NEQ} / \partial I_1^e, \partial \hat{\Phi}_1/\partial I_1^\tau, \cdots$. The resulting ODE trajectories and input-output maps for one model are shown in Fig. \ref{fig02}.

\subsection{Integration of the evolution equation}
We offer two seperate methods for integrating the evolution equations. The first method uses the iterative predictor-corrector algorithm reported in \cite{reeseTheoryFiniteViscoelasticity1998} for the matrix phase and the Newton-Raphson solution scheme reported in \cite{nguyenModelingAnisotropicFinitedeformation2007} for the fiber phase. This solution method is necessary when the deformations are arbitrary. However, the Newton-Raphson algorithms used in this method make it computationally costly and unstable when combined with gradient based optimizers. 

The second method can be used for biaxial experiments, i.e., when plane stress conditions hold and $\F$ is diagonal. Under these circumstances the solution of the evolution equation can be expressed as a system of ODEs which makes convergence much faster and stable.

\subsubsection{Method 1: Predictor-corrector algorithm for arbitrary $\F$}
\subsubsection*{Matrix phase}
The matrix evolution equation provided in \eqref{eq_lie_to_phi} is rewritten here
\begin{align}
    -\frac{1}{2} (\mathscr{L} \be^M) {\b_e^M}^{-1} = \partiald{\Phi}{\tauneq^M} \,.
\end{align}
The key idea is to carry out an operator split of the material time derivative of $\be$ into elastic predictor (E) and inelastic corrector (I) parts
\begin{align}
    \dot{\b}_e^M = \dot{\overline{(\F {\C_i^M}^{-1} \F^T)}} = \underbrace{\mathbf{l}\b_e^M + \b_e^M\mathbf{l}^T}_{\text{E}} + \underbrace{\F \dot{\overline{{\C_i^M}^{-1}}} \F^T}_{I}
\end{align}
where $\mathbf{l} = \Fdot \Finv$ is the spatial velocity gradient. 

In the elastic predictor step the material time derivative of ${\C_i^M}^{-1}$ is set to zero, giving us
\begin{align}
    ({\C_i^M}^{-1})_{\text{trial}} = ({\C_i^M}^{-1})_{t=t_{n-1}} \implies (\be^M)_{\text{trial}} = (\F)_{t=t_{n}} ({\C_i^M}^{-1})_{t=t_{n-1}} (\F)_{t=t_{n}}
\end{align}

In the inelastic corrector step the spatial velocity gradient is set to zero, giving us 
\begin{align}
    -\frac{1}{2} (\mathscr{L} \be^M) {\b_e^M}^{-1} = -\frac{1}{2}\dot{\b}_e^M {\b_e^M}^{-1} &= \partiald{\Phi}{\tauneq^M}
    \\
    \implies \dot{\b}_e^M &= -2\partiald{\Phi}{\tauneq^M} \be^M
\end{align}
Solving this differential equation by exponential mapping and discretizing leads to
\begin{align}
    (\be^M)_{t=t_{n}} \approx \exp \left[ -2 \Delta t \left( \partiald{\Phi}{\tauneq^M} \right)_{t=t_{n}}  \right] (\be^M)_{\text{trial}}
\end{align}
But in the isotropic case $\be^M$ and $\tauneq^M$ are co-axial, which allows us to write this equation in the principal axes
\begin{align}
    {\lambda_{e,i}^M}^2 = \exp \left[ -2\Delta t \, \mathrm{diag} \left(\partiald{\Phi}{\mathbf{\tau}_{i}} \right) \right] ({\lambda_{e,i}^M}^2)_{\text{trial}}
\end{align}
where $\lambda_{e,i}^M$ are the principal elastic stretches of the matrix phase, $\mathrm{diag} \left(\partial {\Phi}/\partial {\mathbf{\tau}_{i}} \right)$ is a diagonal matrix consisting of the derivatives of $\Phi$ with respect to principal values of $\tauneq^M$. Finally, taking the logarithm of both sides gives us
\begin{align}
    \label{eq_epsupdate}
    \epsilon_{e,i} = -\Delta t \, \mathrm{diag} \left(\partiald{\Phi}{\mathbf{\tau}_{i}} \right) +  (\epsilon_{e,i})_{\text{trial}}
\end{align}
The nonlinear equation of \eqref{eq_epsupdate} can be solved using a Newton-Raphson algorithm.

\subsubsection*{Fiber phase}
For an arbitrary $\F$ and corresponding $\C=\F^T\F$, using the scalar split of the fiber stretch in eq. (\ref{eq_scalar_split_F}), the ODE in eq. (\ref{eq_fiber_evol_node}) can be explicitly expressed in terms of $\lambda_i^F$,

\begin{align}
    \label{eq_fiber_evol_ODE}
    \left(\frac{\dot{\lambda}_i^F}{\lambda_i^F} \right) = \mathcal{N}\left(\tau_{NEQ}^F(\lambda_i^F, \mathbf{C}, \mathbf{M}) \right) \, .
\end{align}
Discretizing this differential equation using a backward Euler scheme results in 
\begin{align}
    (\lambda_i^F)_{t_n} = (\lambda_i^F)_{t_{n-1}} + \Delta t (\lambda_i^F)_{t_n} \mathcal{N}\left( \tau_{NEQ}^F((\lambda_i^F)_{t_n}, \mathbf{C}, \mathbf{M}) \right)  \, .
\end{align}
This constitutes a nonlinear and implicit equation for $\lambda_i^F$ at time $t_n$ since $\mathbf{M}$ is constant and $\mathbf{C}$ is prescribed. This equation can be solved for the new value of the internal variable with a numerical solver such as the Newton-Raphson algorithm. The contribution to the stress from the non-equilibrium fiber branch is specified in (\ref{eq_sigma_neq_F}). 

\subsubsection{Method 2: System of ODEs when $\F$ is diagonal and plane stress conditions hold}
\subsubsection*{Matrix phase}
The Lie derivative of $\be^M$ in the evolution equation can be written in terms of $\bedot^M$ to get
\begin{align}
    \label{eq_bedot}
    \bedot^M - (\mathbf{l}  \be^M + \be^M  \mathbf{l}^T) &= -\partiald{\Phi}{\tauneq^M}\be^M
\end{align}
where
\begin{align}
    \mathbf{l} &= \Fdot \Finv
    \\
    \be^M &= \Fe^M {\Fe^M}^T
    \\
    \bedot^M &= \mathbf{\dot{F}}_e^M  (\mathbf{F}_e^M)^T + \mathbf{F}_e ^M (\mathbf{\dot{F}}_e^M)^T
\end{align}
and since $\tauneq^M$ is diagonal, 
\begin{align}
    \partiald{\Phi}{\tauneq^M} = \mathrm{diag} \left(\partiald{\Phi}{\mathbf{\tau}_{i}} \right) \, .
\end{align}
Substituting these back to \eqref{eq_bedot} results in 
\begin{align}
    \label{eq_allFs}
    \Fedot^M  {\Fe^M}^T + \Fe^M  ({\Fedot^M})^T - \left(\Fdot  \Finv  \Fe^M  {\Fe^M}^T + \Fe^M  {\Fe^M}^T  \FinvT \Fdot^T \right) = \mathrm{diag} \left(\partiald{\Phi}{\mathbf{\tau}_{i}} \right)\Fe \Fe^T \, .
\end{align}
where, in the absence of shear terms, every matrix is diagonal, i.e.
\begin{align}
    \F &= \begin{bmatrix} 
    \lambda_1 & 0 & 0 \\
    0 & \lambda_2 & 0 \\
    0 & 0 & \lambda_3
    \end{bmatrix},
    \Fe^M = \begin{bmatrix}
    \lambda_{e,1}^M & 0 & 0 \\
    0 & \lambda_{e,2}^M & 0 \\
    0 & 0 & \lambda_{e,3}^M
    \end{bmatrix},
    \Fdot = \begin{bmatrix} 
    \dot{\lambda}_1 & 0 & 0 \\
    0 & \dot{\lambda}_2 & 0 \\
    0 & 0 & \dot{\lambda}_3
    \end{bmatrix},
    \cdots
\end{align}
Then \eqref{eq_allFs} can be written as a system of equations, which after some simplifications results in 
\begin{align}
    \dot{\lambda}_{e,1}^M &= \left( \frac{\dot{\lambda}_1}{\lambda_1} - \frac{1}{2}\partiald{\Phi}{\tau_{1}} \right) \lambda_{e,1}^M 
    \label{eq_ODE1}
    \\
    \dot{\lambda}_{e,2}^M &= \left( \frac{\dot{\lambda}_2}{\lambda_2} - \frac{1}{2} \partiald{\Phi}{\tau_{2}}\right) \lambda_{e,2}^M 
    \label{eq_ODE2}
    \\
    \dot{\lambda}_{e,3}^M &= \left( \frac{\dot{\lambda}_3}{\lambda_3} - \frac{1}{2} \partiald{\Phi}{\tau_{3}}\right) \lambda_{e,3}^M \, .
    \label{eq_ODE3}
\end{align}

\subsubsection*{Biaxial loading}
Under biaxial loading conditions the out of plane stress component and its time derivative are equal to zero, $\mathbf{\sigma}_{33}^M = \dot{\mathbf{\sigma}}_{33}^M = 0$. Then, assuming the fibers are in the $1-2$ plane,
\begin{align*}
    \dot{\sigma}_{33}^M = \frac{d\sigma_{33}^M}{dt} = \partiald{\sigma_{33}^M}{\lambda_1}\dot{\lambda}_1 + \partiald{\sigma_{33}^M}{\lambda_2}\dot{\lambda}_2 + \partiald{\sigma_{33}^M}{\lambda_3}\dot{\lambda}_3 + \partiald{\sigma_{33}^M}{\lambda_1^e}\dot{\lambda}_{e,1}^M + \partiald{\sigma_{33}^M}{\lambda_2^e}\dot{\lambda}_{e,2}^M + \partiald{\sigma_{33}^M}{\lambda_3^e}\dot{\lambda}_{e,3}^M = 0
\end{align*}
\begin{align}
    \label{lm3dot}
    \implies \dot{\lambda}_3 = A + B\dot{\lambda}_3^e
\end{align}
with
\begin{align*}
    A &= -\frac{\partiald{\sigma_{33}^M}{\lambda_1}\dot{\lambda}_1 + \partiald{\sigma_{33}^M}{\lambda_2}\dot{\lambda}_2 + \partiald{\sigma_{33}^M}{\lambda_{e,1}^M}\dot{\lambda}_{e,1}^M + \partiald{\sigma_{33}^M}{\lambda_{e,2}^M}\dot{\lambda}_{e,2}^M}{\partial \sigma_{33}^M/\partial \lambda_3}
    \\
    B &= -\frac{\partial \sigma_{33}^M/\partial \lambda_3^e}{\partial \sigma_{33}^M/\partial \lambda_3} \, ,
\end{align*}
substituting this into \eqn{eq_ODE3} leads to
\begin{align}
    \label{eq_ODE3_2}
    \dot{\lambda}_{e,3}^M =  \frac{\frac{A}{\lambda_3} - \frac{1}{2} \partiald{\Phi}{\tau_{3}}}{1-\frac{B}{\lambda_3}\lambda_{e,3}^M} \lambda_{e,3}^M
\end{align}
Then equations \eqref{eq_ODE1}, \eqref{eq_ODE2} and \eqref{eq_ODE3_2} constitute a system of three ODEs for the three unknowns $\lambda_{e,1}^M, \lambda_{e,2}^M$ and $\lambda_{e,3}^M$.

\subsubsection*{Uniaxial loading}
The case of uniaxial loading closely resembles the case of biaxial loading. The only difference being the symmetry $\lambda_2=\lambda_3$, $\dot{\lambda}_2=\dot{\lambda}_3, \cdots$. This simplifies the equations \eqref{eq_ODE1}, \eqref{eq_ODE2} and \eqref{eq_ODE3_2} to 
\begin{align}
    \dot{\lambda}_{e,1}^M &= \left( \frac{\dot{\lambda}_1}{\lambda_1} - \frac{1}{2}\partiald{\Phi}{\tau_{11}} \right) \lambda_1^e 
    \\
    \dot{\lambda}_{e,2}^M &= \dot{\lambda}_3^e
    \\
    \dot{\lambda}_{e,3}^M &=  \frac{\frac{A'}{\lambda_3} - \frac{1}{2} \partiald{\Phi}{\tau_{33}}}{1-\frac{B'}{\lambda_3}\lambda_{e,3}^M} \lambda_{e,3}^M
\end{align}
with
\begin{align*}
    A' &= -\frac{\partiald{\sigma_{33}^M}{\lambda_1}\dot{\lambda}_1 + \partiald{\sigma_{33}^M}{\lambda_{e,1}^M}\dot{\lambda}_{e,1}^M}{\partial \sigma_{33}^M/\partial \lambda_2 + \partial \sigma_{33}^M/\partial \lambda_3}
    \\
    B' &= -\frac{\partial \sigma_{33}^M/\partial \lambda_{e,2}^M + \partial \sigma_{33}^M/\partial \lambda_{e,3}^M}{\partial \sigma_{33}^M/\partial \lambda_2 + \partial \sigma_{33}^M/\partial \lambda_3} \, ,
\end{align*}

\subsubsection*{Fiber phase}

Note that the fiber evolution equation of \eqref{eq_fiber_evol_ODE} is an ordinary differential equation in $\lambda_i^F$. This equation can be integrated using a numerical integrator of choice thus avoiding an iterative solution scheme. This integration can be performed independent of the matrix phase in biaxial experiments if the fibers are assumed to lie in the $1-2$ plane.

\subsection{Model calibration and verification}
We use both experimental and synthetic training data to train and test our models. The synthetic data is generated using the creep potential used in \cite{reeseTheoryFiniteViscoelasticity1998} and neo-Hookean material models for the Helmholtz free energy functions. The creep potential in \cite{reeseTheoryFiniteViscoelasticity1998} can be shown to have the following form
\begin{align}
    \label{eq_RG_Phi}
    \Phi_{RG} &= \frac{1}{9\eta_V}(I_1^\tau)^2 + \frac{1}{3\eta_D}((I_1^\tau)^2-3I_2^\tau) \, .
\end{align}
Note that differentiating this form of the creep potential with respect to $\tauneq$ twice leads to the familiar isotropic matrix $\mathcal{V}^{-1} = \frac{1}{2\eta_D} (\mathbb{I} - \frac{1}{3} \mathbf{I} \otimes \mathbf{I}) + \frac{1}{9\eta_V} \mathbf{I} \otimes \mathbf{I}$ reported in \cite{reeseTheoryFiniteViscoelasticity1998}, where $\eta_D$ and $\eta_V$ are material parameters and $\mathbb{I}$ is the 4\ss{th} order identity tensor.

In the case of anisotropic data we use a linear form of the evolution equation \eqref{eq_diss_ineq_fiber} as proposed in \cite{nguyenModelingAnisotropicFinitedeformation2007}
\begin{align}
    \left( \frac{\dot{\lambda}_i^F}{\lambda_i^F} \right) = \frac{1}{\eta_F} \tau_{NEQ}^F
\end{align}
where $\eta_F$ is a positive parameter that represents the characteristics viscosity of the fiber.

We use the following form of neo-Hookean strain energy functions for $\Psi^M_{EQ}$ and $\Psi^M_{NEQ}$ when generating synthetic training data
\begin{align}
    \label{eq_neoHook}
    \Psi = \frac{1}{2}\mu(J^{-2/3}\tr{\b}-3) + \frac{1}{2}K(J-1)^2
\end{align}
and the following form for $\Psi^F_{EQ}$ and $\Psi^F_{NEQ}$
\begin{align}
    \Psi = k(\exp(I_4-1)-1) \, .
\end{align}

We use experimental data obtained from biaxial stress relaxation experiments on blood clots \cite{rauschHyperviscoelasticDamageModeling2021}, natural rubber \cite{aminImprovedHyperelasticityRelation2002}, four brain tissue samples \cite{buddayRheologicalCharacterizationHuman2017} and myocardium \cite{sommerBiomechanicalPropertiesMicrostructure2015} to train our models.

\section{Results}
\begin{figure*}[h!]
\centering
\includegraphics{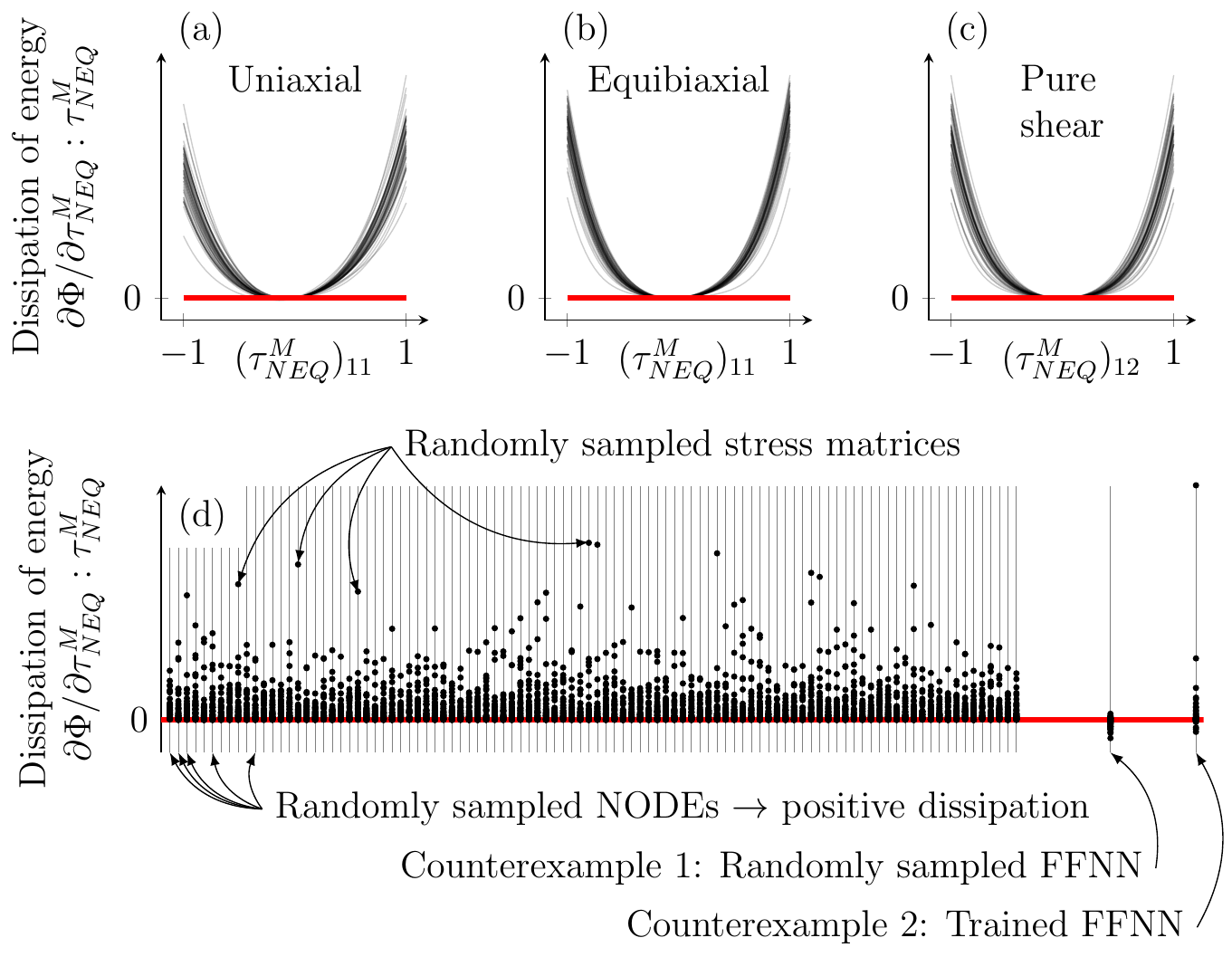} \vspace{-4mm}
\caption{The NODE modeling framework is guaranteed to result in non-negative dissipation of energy under all deformations. Dissipation of energy under uniaxial, equibiaxial and pure shear conditions with 50 randomly sampled NODEs (a-c), and, dissipation of energy under 50 randomly sampled states of stress ($\tauneq$) with 100 randomly sampled NODEs, 1 randomly sampled FFNN and 1 trained FFNN on physically admissible data (d). All points above the red lines correspond to positive dissipation of energy.}
\label{fig03} 
\end{figure*}

A physically sound formulation of viscoelasticity satisfies the 2\ss{nd} law of thermodynamics. To this end, we design creep potentials $\Phi$ with appropriate requirements to satisfy this criterion. A demonstration of this is shown in Fig. \ref{fig03} where, even randomly sampled NODEs are shown to result in non-negative dissipation of energy under arbitrary stresses. On the other hand, feed-forward neural networks (FFNN) fail to satisfy this criterion. We show the dissipation of energy with two FFNN-based creep potentials, one randomly sampled like the NODEs and the other trained against the creep potential of \eqref{eq_RG_Phi}. We expect that the randomly sampled FFNN would violate the 2\ss{nd} law of thermodynamics, but even the FFNN trained to interpolate an appropriate creep potential violates this criterion especially outside the training region.

\begin{figure*}[h!]
\centering
\includegraphics{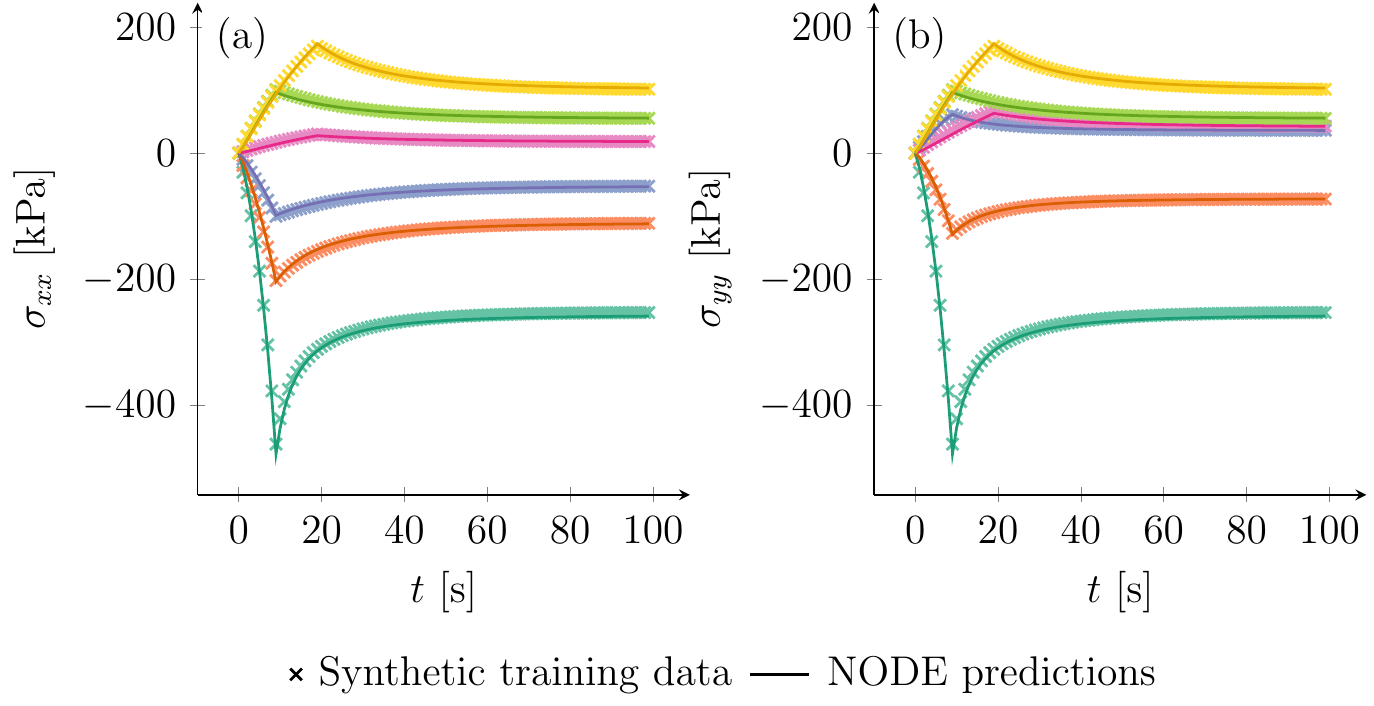}
\caption{Results of training the NODEs against synthetic data generating with the analytical model described in \eqref{eq_RG_Phi} and \eqref{eq_neoHook}.}
\label{fig04} 
\end{figure*}

We design the creep potential with a very expansive basis to enhance the capability of our model to capture all modes of energy dissipation. In particular, our formulation encompasses a generalization of the creep potentials of the form \eqref{eq_RG_Phi} which are ubiquitously used to describe viscoelastic behavior of isotropic materials. To demonstrate this we generate synthetic stress relaxation data with the creep potential of \eqref{eq_RG_Phi} and neo-Hookean models and train the NODE based model against this data. The resulting plots of stress for various rates of loading and peak stress are shown in Fig. \ref{fig04}. As expected, the NODE based model learns and replicates the training data almost exactly. 

Next we train the model with experimental data. Training a data-driven model of viscoelasticity is computationally expensive. This is due to the fact that the state of stress at any given point depends not only on the observable quantities like the strains, but also on the values of the hidden variables such as $\lambda_{e,1}^M, \lambda_{e,2}^M$, $\lambda_{e,3}^M$ and $\lambda_i^F$ in biaxial experiments, or $({\C_i^M})^{-1}$, $\lambda_i^F$ in the case of arbitrary deformations. This makes training data-driven models time consuming as it entails a full integration of the entire history of stress in every epoch of training. Therefore, we rely on closed-form models of the form \eqref{eq_RG_Phi} and \eqref{eq_neoHook} to \emph{pre-train} the NODE based model. This reduces training time because in this case the potentials governing the stresses, $\Phi$, $\Psi^M_{EQ}, \Psi^M_{NEQ}, \Psi^F_{EQ}$ and $\Psi^F_{NEQ}$ can be trained individually without integration of stress. This gives us a rough approximation of the parameters. Then the NODEs can be \emph{re-trained} against the actual stress data directly to pinpoint the parameters of the model.

\begin{figure*}[h!]
\centering
\includegraphics{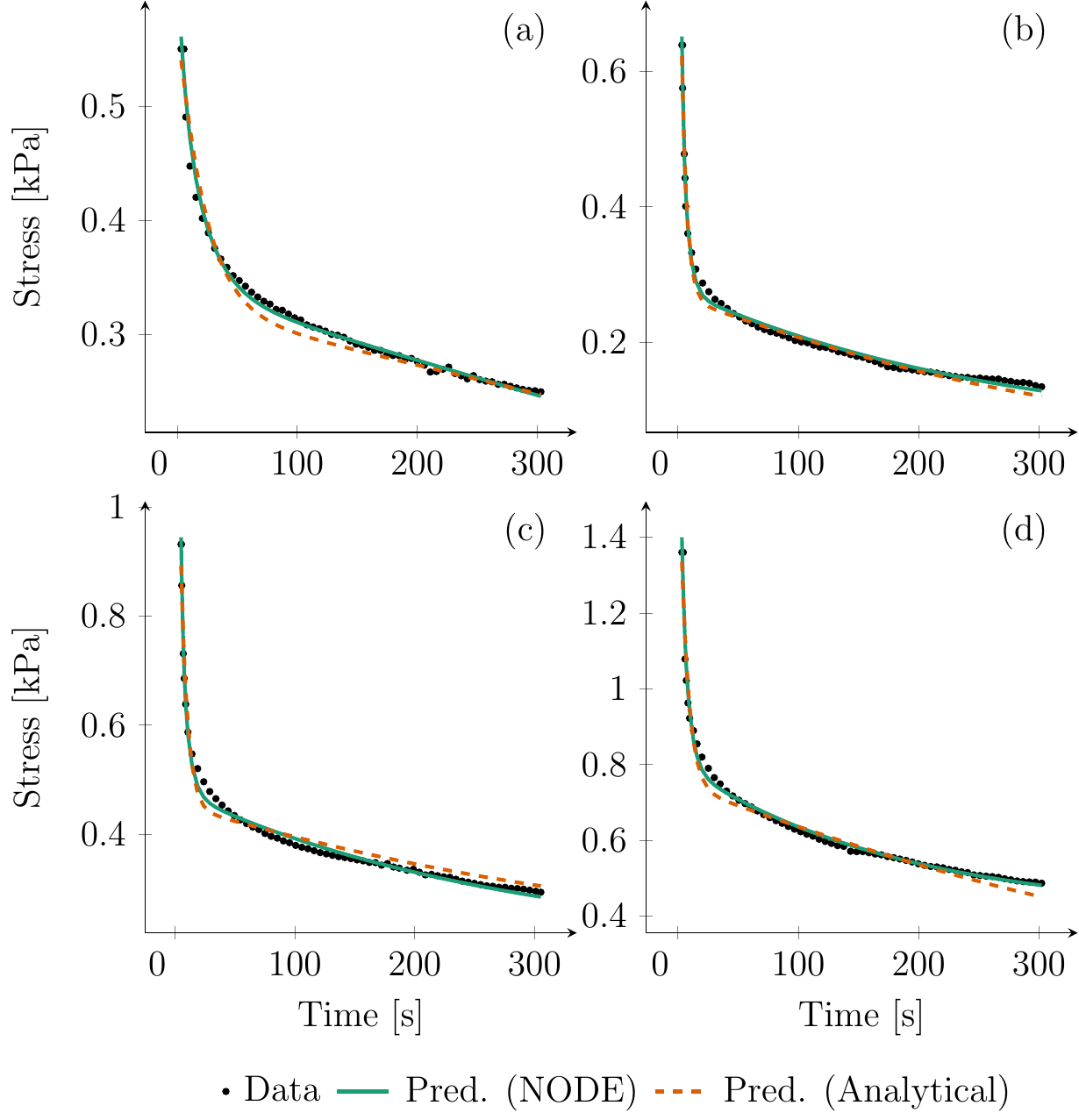}
\caption{Compression relaxation data for tissue samples obtained from various regions of brain and predictions of the NODE and the analytical model after training on each dataset. From top left clockwise: Basal ganglia (a), Carpus callosum (b), Corona radiata (c), Cortex (d).}
\label{fig05} 
\end{figure*}

As a first step we train the model with experimental stress-relaxation data from human brain tissue \cite{buddayRheologicalCharacterizationHuman2017}. The training data consists of compression relaxation experiments on four regions of the brain; basal ganglia, carpus callosum, corona radiata and cortex. The plots of the data and the predictions of our model after training are shown in Fig. \ref{fig05}. We train the analytical model to benchmark the performance of the NODEs. On average, the NODE based model has a mean absolute error (MAE) of 7.3 Pa, whereas the analytical model has a nearly 80\% higher error at 13.3 Pa. 

\begin{figure*}[h!]
\centering
\includegraphics{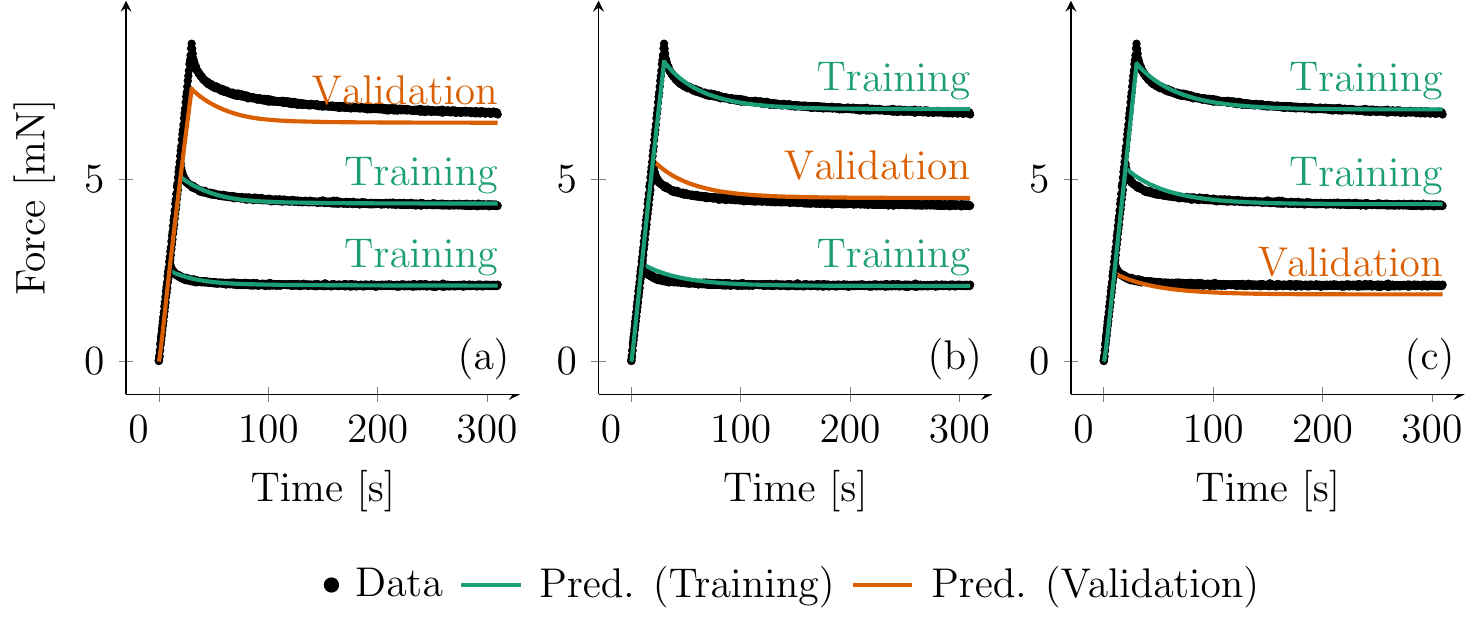}
\caption{Stress relaxation data obtained from blood clots and predictions of the model after training.}
\label{fig06} 
\end{figure*}

Next, we train the model with data on blood clots \cite{rauschHyperviscoelasticDamageModeling2021}. This dataset consists of three stress relaxation experiments with on a blood clot sample with peak stresses of 10\%, 20\% and 30\%, respectively. In the experiments a $10\text{mm} \times 40\text{mm}$ blood clot sample with a thickness of 2 mm is stretched along the short edge with a strain rate of 0.001 m/s up to the desired peak stress and held for 300 s. We train with two of the datasets at a time while testing with the third experiment. The data and predictions of the models after training are shown in Fig. \ref{fig06}. In Fig. \ref{fig06} (a) the training data consists of the two curves with the lower peak force, in Fig. \ref{fig06} (b) it consists of the lower and the upper curves while the middle curve is used for validation and in Fig. \ref{fig06} (c) the lowermost curve is kept for validation while the upper two curves are used for training. As we go from Fig. \ref{fig06} (a) to (c) the training MAE steadily increases from 0.033 mN to 0.056 mN to 0.070 mN. This is partly because the magnitude of training stress is higher as we go from (a) to (c). The rise in \emph{mean relative error} (MRE) is much less pronounced at 0.015, 0.019 and 0.021, respectively. The validation MAE, on the other hand, is lowest for Fig. \ref{fig06} (b) at 0.193 mN compared to MAEs of 0.437 mN and 0.209 mN of the other two cases. This indicates that the model performs better in interpolation than extrapolation. Nevertheless, the prediction in the validation dataset still captures qualitatively and quantitatively the response of the blood clot. 

\begin{figure*}[h!]
\centering
\includegraphics{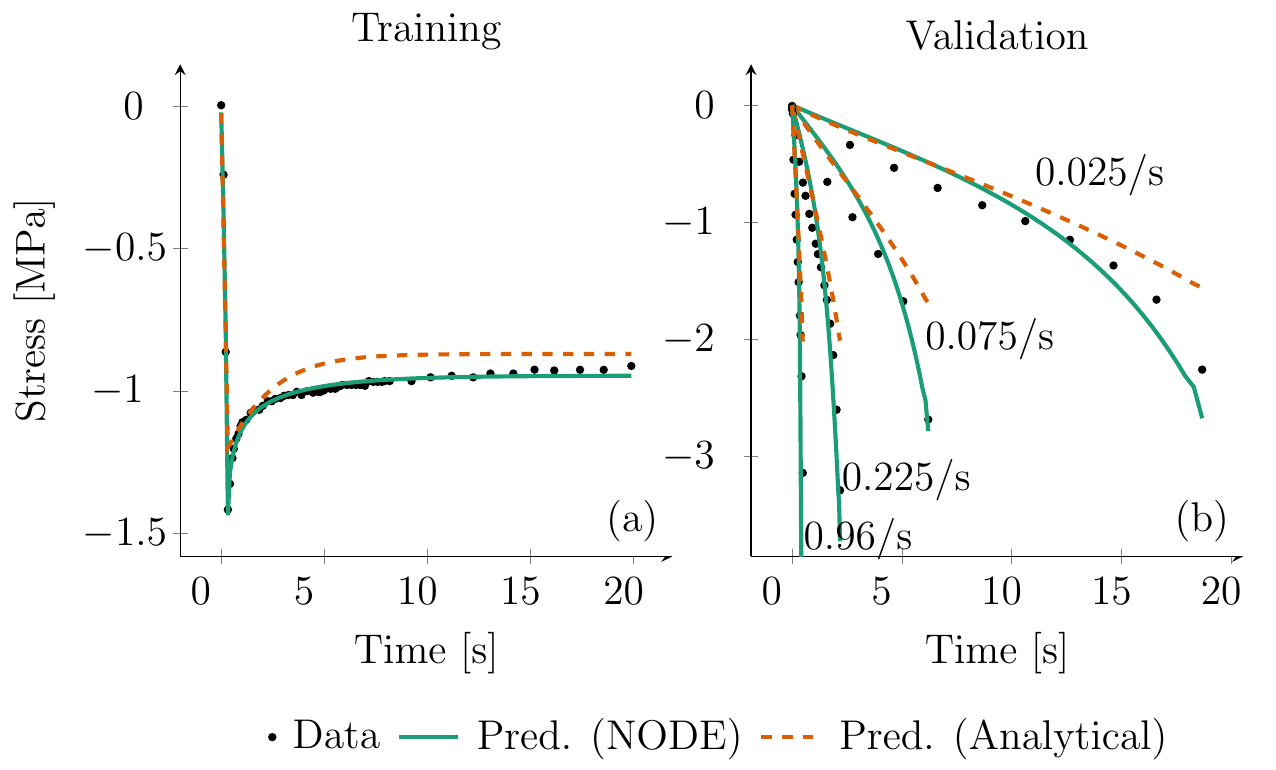}
\caption{The NODE and the analytical model were trained with experimental stress relaxation data on natural rubber (a) and tested with various monotone compression curves with different strain rates (b).}
\label{fig07} 
\end{figure*}

Lastly, we train the isotropic model with data from natural rubber \cite{aminImprovedHyperelasticityRelation2002}. The data for natural rubber consists of both a stress relaxation and various monotonic compression experiments under different strain rates. We train the model with the stress relaxation data and test with the monotonic compression curves as a validation. The results are shown in Fig. \ref{fig07}. The plots of Fig. \ref{fig07} (a) show that the NODE captures the training data almost exactly, while the analytical model struggles to do so. The MAE of the analytical model is approximately 5 times higher than that of the NODE at 0.071 MPa for the analytical model versus 0.014 MPa of the NODE model. Qualitatively, there is a notable difference in the analytical compared to the NODE predictions in the Training plot of Fig. \ref{fig07}. The analytical model is simply incapable of capturing the response, but this is not a problem for the NODE model. 

In validation, NODE has lower errors in the three curves with 0.025/s, 0.075 and 0.225/s strain rates with an MAE between 0.131 and 0.231 MPa whereas the MAE of the analytical model ranges from 0.218 to 0.351 MPa, a much poorer performance. On the 0.96/s curve the analytical model has the lower error at 0.413 MPa vs 0.672 MPa of the NODE. This demonstrates that the model yields acceptably close results even when trained with one type of testing data and tested with another.

\begin{figure*}[h!]
\centering
\includegraphics{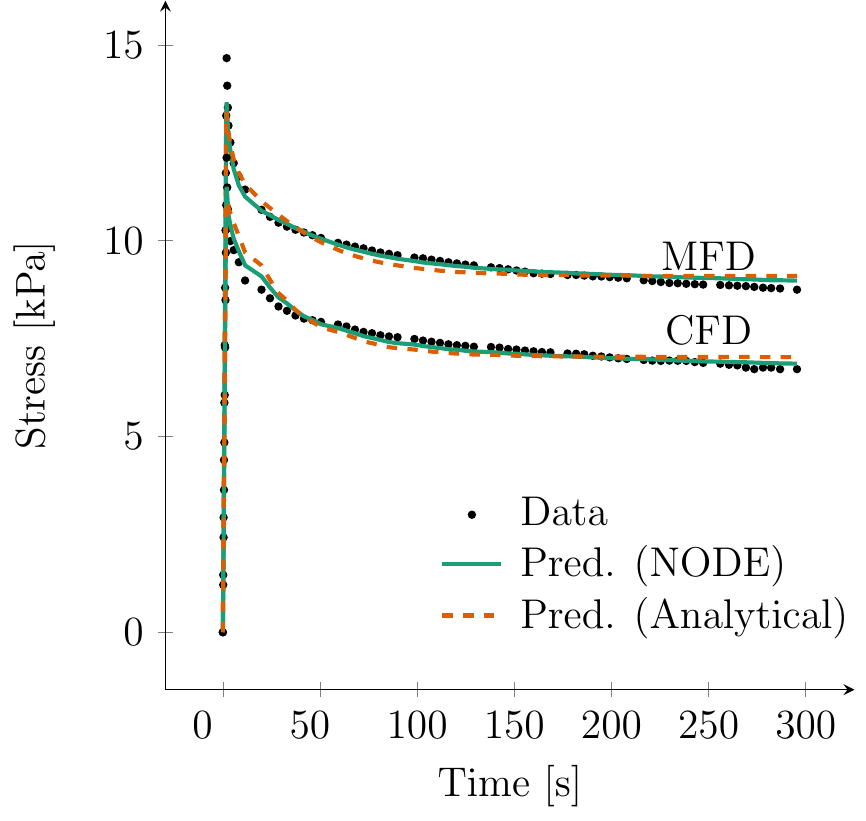}
\caption{Stress relaxation data for myocardium in Mean Fiber Direction (MFD) and Cross Fiber Direction (CFD) and the predictions of the two models after training.}
\label{fig08} 
\end{figure*}

Next, we train the model with anisotropic data. For this we use stress relaxation data for human myocardium \cite{sommerBiomechanicalPropertiesMicrostructure2015}. The data consists of the average stress of $n=5$ human myocardium samples in the Mean Fiber Direction (MFD) and Cross Fiber Direction (CFD) in a 5 min long stress relaxation experiment with a 10\% peak stress. The results of training both analytical and NODE-based models are shown in Fig. \ref{fig08}. The NODE has MAEs of 0.187 kPa and 0.145 kPa in MFD and CFD, respectively, whereas the analytical model has MAEs of 0.327 kPa and 0.271 kPa.

A summary of the performance of the two models under all experimental training data is provided in Fig. \ref{fig08}. We use mean relative error in this figure to unify the vertical axis of all the different materials. The NODE has the lower error in \emph{all} cases demonstrating the flexibility of the data-driven model.

\begin{figure*}[h!]
\centering
\includegraphics{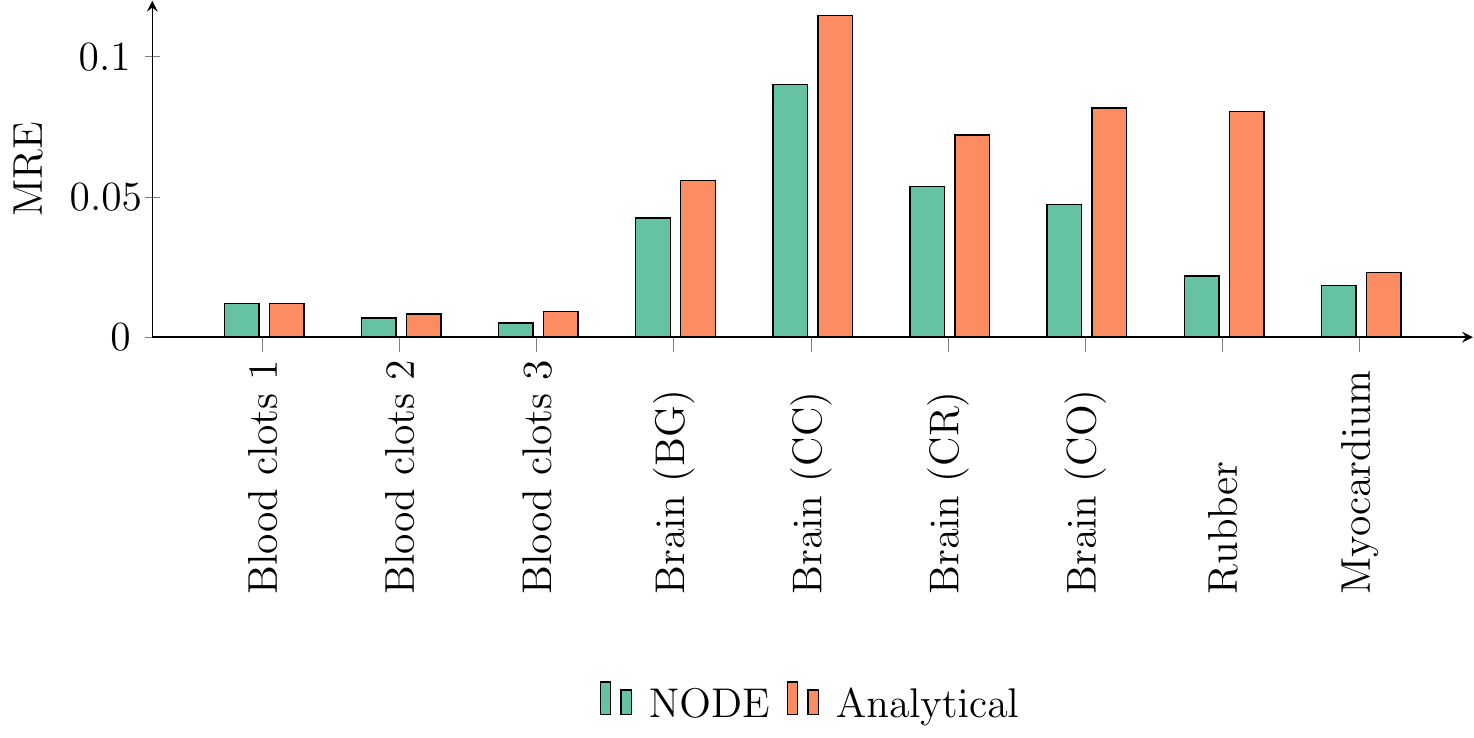}
\caption{Training mean relative error (MRE) for all experimental data considered in this study for the NODE and the analytical model.}
\label{fig09} 
\end{figure*}

\section{Discussion}
We present a fully data-driven, physics-informed model of finite viscoelasticity. Unlike theories of linear or finite linear viscoelasticity, finite viscoelasticity allows for large deformations and large deviations from the thermodynamic equilibrium \cite{reeseTheoryFiniteViscoelasticity1998}. This functionality is especially important for soft materials like rubber and soft biological tissue because such materials routinely undergo extremely large deformations in dynamic environments. 

The state of stress in a viscoelastic solid can be described using two class of potentials, a creep or dissipation potential and the Helmholtz free energy function \cite{kumar2016two}. In the classical approach to viscoelasticity both of these potentials are modeled with closed-form expressions. The creep potential of eq. \eqref{eq_RG_Phi} is nearly the \emph{default} isotropic creep potential of choice for most studies. Some examples of its use for modeling various materials like blood clots, polymers and rubber can be found in \cite{rauschHyperviscoelasticDamageModeling2021, kleuterGeneralizedParameterIdentification2007, loewRatedependentPhasefieldDamage2019}. Some other studies extend the model of \cite{reeseTheoryFiniteViscoelasticity1998} to the anisotropic case using an isotropic-anisotropic split of the Helmholtz free energy function \cite{liuAnisotropicFiniteStrain2019, nguyenModelingAnisotropicFinitedeformation2007}, but even in these cases, linear dissipation is used for the isotropic and anisotropic parts. 

The dissipation potential eq. \eqref{eq_RG_Phi} is a simple function with only two parameters. While it has remained popular, the restrictions of a linear model limit the flexibility of viscoelastic materials that can be modeled. Nonlinear extensions of eq. \eqref{eq_RG_Phi} in the literature retain the same functional form but replace the viscosity with a function of the deformation \cite{amin2006nonlinear,zhou2018micro}. We offer a physics-informed, data-driven potential based on NODEs that has much more flexibility, allowing it to capture complex viscoelastic dissipation patterns. In fact, it can be shown that the dissipation potential of \eqref{eq_RG_Phi} is a special case of the proposed potential when the outputs of the NODEs are set to constant values rather than convex functions of their arguments. 

The methodology proposed in this paper encompasses a large class of viscoelastic dissipation response. We require that the dissipation potential be a convex function with its global minimum at $\tauneq^M=0$ as stated in Section \ref{theory_of_visco}. However, this is not necessary, as \emph{all} functions 
 $f$ whose partial derivative $\partial {f}/ \partial {(\tauneq^M)_{ij}}$ with respect to the elements $(\tauneq^M)_{ij}$ of $\tauneq^M$ is positive when $(\tauneq^M)_{ij}\geq 0$ and negative when $(\tauneq^M)_{ij}\leq 0$ will satisfy the positive dissipation criterion. Nevertheless, convexity is an elegant and mathematically convenient way of expressing functions that satisfy this criterion \cite{kumar2016two}. Furthermore, we choose to work with only three functions of the invariants of $\tauneq$ ($g_1, g_2$ and $g_3$), but there are more possibilities that could be utilized to build even wider classes of constitutive models. Practically speaking, even with just three invariants of $\tauneq$ we can already extend the class of viscoelastic behavior way beyond the linear potentian Eq.  \eqref{eq_RG_Phi}, and beyond the nonlinear extensions which keep Eq.  \eqref{eq_RG_Phi} but turn the viscosity into a function of the deformation \cite{zhou2018micro}. The NODE functions can capture arbitrary convex functions (provided enough neurons are used to represent the creep potentials). When trained on various experimental data, the NODE has no problem fitting the data, see Figs. (\ref{fig05}-\ref{fig08}). 

 Other theories to model viscoelasticity, in particular for polymers, involves the use of statistical arguments to decribe the slip and reorganization of the polymer chains \cite{vernerey2017statistically}. A similar approach, based on reactive mixtures, has been proposed for biological tissue \cite{ateshian2015viscoelasticity}. Another large class of models can be derived from extensions of the linear viscoelasticity models to large deformations. For example, in \cite{holzapfel1996new}, non-equilibrium stresses are introduced as internal variables, and each satisfies a linear ordinary differential equation. This class of models result in sums of exponential relaxation functions. To capture non-exponential decay, extensions of \cite{holzapfel1996new} using fractional derivatives can result in richer relaxation behavior \cite{zhang2020efficient,freed2006fractional}.

 Regarding the Helmholtz free energy, we have previously demonstrated that NODEs can be used to construct polyconvex and objective strain energy functions for modeling hyperelastic materials \cite{tacDatadrivenTissueMechanics2022}. Therefore, for the equilibrium potential in the present work we use almost the same methodology as in \cite{tacDatadrivenTissueMechanics2022}. One change in the present formulation is the modeling of compressible and nearly incompressible behavior as opposed to fully incompressible materials. The non-equilibrium free energy function has the same requirements as the equilibrium potential, only with respect to the elastic component of the deformation. Thus, the NODE framework can be used as well for the calculation of the non-equilibrium stresses. Taken together, the proposed methodology is able to describe the entire material behavior using data-driven potentials, without the need of a single \textit{ad hoc} material parameter or constitutive relation.


 The flexibility of the method is demonstrated by its ability to learn the viscous behavior of a wide range of biological and synthetic materials including multiple brain tissue samples, rubber, blood clots, and human myocardium. The training error in all cases is consistently lower than the analytical model of \cite{reeseTheoryFiniteViscoelasticity1998}. The results of training with data from natural rubber also indicates that the model is capable of predicting stresses in \emph{unseen} types of loading. The framework is suitable for isotropic or anisotropic materials. Importantly, all this flexibility is achieved while polyconvexity and positive energy dissipation are satisfied exactly.
 
\section{Conclusions}
In this paper we introduce a physics-informed data-driven model of general, finite, anisotropic, three-dimensional viscoelasticity based on NODEs. We show that the dissipation potential based on NODEs automatically satisfies the 2\ss{nd} law of Thermodynamics by leveraging properties of NODEs to build monotonic functions.  Polyconvexity of the Helmholtz free energy is also ensured with a NODE framework. We train the proposed model with various experimental datasets and demonstrate the flexibility of the method in capturing complex viscoelastic responses of a wide class of materials including rubbers, blood clots, brain, and myocardium. We anticipate that this work will enable the modeling and simulation of arbitrary viscoelastic materials without the burden of selecting closed-form constitutive equations. 

\section{Acknowledgements}
This work was supported by NIAMS award R01AR074525, NSF CMMI award 1916668.

\section{Supplementary material}
All data, model parameters and code associated with this study are available in a public Github repository at \url{https://github.com/tajtac/nvisco}.

\appendix
\section{Proof of convexity of $I_1^2$ and $I_1^2-\alpha I_2$}
\label{ap_proof_convex}
A function $f$ of a symmetric $3 \times 3$ matrix $\mathbf{M}$ can alternatively be expressed as a function of the 6 independent elements of $\mathbf{M}$ as
\begin{align}
    f(\mathbf{M}) = f(M_{11}, M_{22}, M_{33}, M_{12}, M_{13}, M_{23}) \, .
\end{align}
The function $f$ is convex in the elements of $\mathbf{M}$ if the Hessian matrix of $f$ is positive semi-definite. The Hessian matrix of $f$ is given by
\begin{align}
    \mathbf{H} = 
    \begin{bmatrix}
        \partiald{^2f}{M_{11}^2} & \partiald{^2f}{M_{11} \partial M_{22}} & \partiald{^2f}{M_{11} \partial M_{33}} & \partiald{^2f}{M_{11} \partial M_{12}} & \cdots
        \\
        \partiald{^2f}{M_{22} \partial M_{11}} & \partiald{^2f}{M_{22}^2} & \partiald{^2f}{M_{22} \partial M_{33}} & \partiald{^2f}{M_{22} \partial M_{12}} & \cdots
        \\
        \partiald{^2f}{M_{33} \partial M_{11}} & \partiald{^2f}{M_{33} \partial M_{22}} & \partiald{^2f}{M_{33}^2} & \partiald{^2f}{M_{33} \partial M_{12}} & \cdots
        \\
        \vdots & \vdots & \vdots & \vdots
    \end{bmatrix}
\end{align}
$\mathbf{H}$ is positive semi-definite if all of its eigenvalues are non-negative. 
\subsection{$I_1^2$}
The function $f(\mathbf{M}) = \left(I_1(\mathbf{M})\right)^2 = (\tr \mathbf{M})^2$ can be expressed as
\begin{align}
    f(\mathbf{M}) = f(M_{11}, M_{22}, M_{33}) = (M_{11} + M_{22} + M_{33})^2
\end{align}
Then it is readily seen that the Hessian of $f$ is
\begin{align}
    \mathbf{H} =
    \begin{bmatrix}
        2 & 2 & 2 & 0 & 0 & 0
        \\
        2 & 2 & 2 & 0 & 0 & 0
        \\
        2 & 2 & 2 & 0 & 0 & 0
        \\
        0 & 0 & 0 & 0 & 0 & 0
        \\
        0 & 0 & 0 & 0 & 0 & 0
        \\
        0 & 0 & 0 & 0 & 0 & 0
    \end{bmatrix}
\end{align}
which has eigenvalues $\lambda_1 = 6, \lambda_2 = \lambda_3 = \lambda_4 = \lambda_5 = \lambda_6 = 0$. Therefore $I_1^2$ is convex.

\subsection{$I_1^2-\alpha I_2$}
The second principal invariant $I_2$ of a matrix $\mathbf{M}$ can be expressed in terms of the elements of $\mathbf{M}$ as
\begin{align}
    I_2(\mathbf{M}) = M_{11}M_{22} + M_{11}M_{33} + M_{22}M_{33} - M_{12}^2 - M_{13}^2 - M_{23}^2
\end{align}
using this in $f(\mathbf{M}) = I_1^2 - \alpha I_2$ results in
\begin{align*}
    f(\mathbf{M}) &= (M_{11} + M_{22} + M_{33})^2 
    \\
    &\phantom{=} - \alpha (M_{11}M_{22} + M_{11}M_{33} + M_{22}M_{33} - M_{12}^2 - M_{13}^2 - M_{23}^2 ) 
    \\
    &= M_{11}^2 + M_{22}^2 + M_{33}^2 + (2-\alpha)(M_{11}M_{22} + M_{11}M_{33} + M_{22}M_{33}) 
    \\
    &\phantom{=} + \alpha (M_{12}^2 + M_{13}^2 + M_{23}^2) \, .
\end{align*}
Then the Hessian of $f$ is given as
\begin{align}
    \mathbf{H} =
    \begin{bmatrix}
        2 & 2-\alpha & 2-\alpha & 0 & 0 & 0
        \\
        2-\alpha & 2 & 2-\alpha & 0 & 0 & 0
        \\
        2-\alpha & 2-\alpha & 2 & 0 & 0 & 0
        \\
        0 & 0 & 0 & 2\alpha & 0 & 0
        \\
        0 & 0 & 0 & 0 & 2\alpha & 0
        \\
        0 & 0 & 0 & 0 & 0 & 2\alpha
    \end{bmatrix} \, .
\end{align}
This matrix has eigenvalues 
\begin{align}
    \lambda_1 &= 2(3-\alpha)
    \\
    \lambda_2 &= \alpha
    \\
    \lambda_3 &= \alpha
    \\
    \lambda_4 &= 2\alpha
    \\
    \lambda_5 &= 2\alpha
    \\
    \lambda_6 &= 2\alpha \, .
\end{align}
Then it can be seen that all the eigenvalues are non-negative for $0 \leq \alpha \leq 3$. Therefore $I_1^2-3I_2$ is convex for $0 \leq \alpha \leq 3$.

\end{document}